\documentclass[12pt]{article}
\usepackage{graphicx}
\usepackage{amssymb}
\usepackage{amsmath}
\usepackage{epsf}

\input{epsf}
\begin{document}
\addtolength{\baselineskip}{.5mm}
\newlength{\extraspace}
\setlength{\extraspace}{1.5mm}
\newlength{\extraspaces}
\setlength{\extraspaces}{2mm}
\newcommand{\figref}[1]{\protect\ref{#1}}

\makeatletter\@addtoreset{equation}{section}\makeatother\renewcommand{\theequation}{\thesection.\arabic{equation}}

\newcommand{\newsection}[1]{
\vspace{15mm}
\pagebreak[3]
\addtocounter{section}{1}
\setcounter{subsection}{0}
\setcounter{footnote}{0}
\noindent
{\Large\bf \thesection. #1}
\nopagebreak
\medskip
\nopagebreak}

\newcommand{\newsubsection}[1]{
\vspace{1cm}
\pagebreak[3]
\addtocounter{subsection}{1}
\addcontentsline{toc}{subsection}{\protect
\numberline{\arabic{section}.\arabic{subsection}}{#1}}
\noindent{\large\bf 
\thesubsection.
#1}
\nopagebreak
\vspace{2mm}
\nopagebreak}
\newcommand{\ba}{\begin{eqnarray}
\addtolength{\abovedisplayskip}{\extraspaces}
\addtolength{\belowdisplayskip}{\extraspaces}
\addtolength{\abovedisplayshortskip}{\extraspace}
\addtolength{\belowdisplayshortskip}{\extraspace}}
\newcommand{\zbar}{\overline{z}}
\newcommand{\tr}{{\rm tr\,}}
\newcommand{\ea}{\end{eqnarray}}
\newcommand{\OL}[1]{ \hspace{2pt}\overline{\hspace{-2pt}#1
   \hspace{-1pt}}\hspace{1pt} }
\newcommand{\is}{& \! = \! &}
\newcommand{\eps}{\epsilon}
\newcommand{\calR}{{\cal R}}
\newcommand{\calB}{{\cal M}}
\newcommand{\calP}{{\cal P}}
\newcommand{\M}{{\cal M}}
\newcommand{\Mv}{{\cal M}_{vector}}
\newcommand{\Mh}{{\cal M}_{hyper}}
\newcommand{\calK}{{\cal K}}
\newcommand{\K}{{\cal K}}
\newcommand{\calG}{{\cal G}}
\newcommand{\G}{{\cal G}}
\newcommand{\N}{{\cal N}}
\newcommand{\F}{{\cal F}}
\newcommand{\tilF}{{\tilde F}}
\newcommand{\tomega}{{\tilde \omega}}
\newcommand{\txi}{{\tilde \xi}}
\newcommand{\hphi}{{\hat \phi}}
\newcommand{\barG}{{\OL G}}
\newcommand{\alphap}{}
\newcommand{\IZ}{\mathbb{Z}}
\newcommand{\IP}{\mathbb{P}}
\newcommand{\mv}{\M_V}
\newcommand{\mh}{\M_H}
\newcommand{\be}{\begin{equation}
\addtolength{\abovedisplayskip}{\extraspaces}
\addtolength{\belowdisplayskip}{\extraspaces}
\addtolength{\abovedisplayshortskip}{\extraspace}
\addtolength{\belowdisplayshortskip}{\extraspace}}
\newcommand{\ee}{\end{equation}}
\newcommand{\STr}{{\rm STr}}
\newcommand{\figuur}[3]{
\begin{figure}[t]\begin{center}
\leavevmode\hbox{\epsfxsize=#2 \epsffile{#1.eps}}\\[3mm]
\parbox{15.5cm}{\small
\it #3}
\end{center}
\end{figure}}
\newcommand{\re}{{\rm Re \,}}
\newcommand{\im}{{\rm Im \,}}
\newcommand{\calm}{{\cal M}}
\newcommand{\sect}[1]{\section{#1}}
\newcommand\hi{{\rm i}}
\newcommand{\ini}{\im \N^{-1}}
\newcommand{\tp}{\tilde{p}}
\newcommand{\tq}{\tilde{q}}

\begin{titlepage}
\begin{center}
{\hbox to\hsize{ \hfill SLAC-PUB-10868}} {\hbox to\hsize{ \hfill SU-ITP-04/42}}

{\hbox to\hsize{ \hfill hep-th/0411279}}

\vspace{3.5cm}

{\Large \bf Moduli Potentials in Type IIA \\[6mm]
Compactifications with RR and NS Flux}
\\[1.5cm]

{Shamit Kachru and Amir-Kian Kashani-Poor}
\\[8mm]

{\it Department of Physics and SLAC, Stanford University,
Stanford, CA 94305/94309}

\vspace*{1.5cm}

{\bf Abstract}\\

\end{center}
\noindent
We describe a simple class of type IIA string compactifications on
Calabi-Yau manifolds where background fluxes
generate a potential for the complex structure moduli, the
dilaton, and the K\"ahler moduli.
This class of models corresponds to gauged ${\cal N}=2$ supergravities,
and the potential is completely determined by a choice
of gauging and by data of the ${\cal N}=2$
Calabi-Yau model -- the prepotential for vector multiplets and
the quaternionic metric on the hypermultiplet moduli space.
Using mirror symmetry,
one can determine many (though not all) of the quantum corrections
which are relevant in these models.

\end{titlepage}

\newcommand{\Tr}{{\rm Tr}}

\newpage
\section{Introduction}

Finding compactifications with computable potentials for the scalar moduli is an important problem in string theory.
Early ideas in this direction, in the context of the heterotic
string, can be found in
\cite{oldhet1,oldhet2}.
More recently, there has been a great deal of activity exploring the
potentials generated by p-form fluxes in type II
string models (for some excellent reviews with references,
see e.g. \cite{Eva,reviews}).  Much of the attention has been focused on
type IIB Calabi-Yau flux vacua,
where the flux-induced potentials
depend on the complex and dilaton moduli, but not the
K\"ahler moduli \cite{IIBflux,GKP}.  In this type IIB context,
there is increasingly strong evidence that proper
incorporation of quantum corrections
(to the superpotential and/or K\"ahler potential)
yields large numbers of models where
the K\"ahler moduli can be stabilized as well \cite{KKLT,DDF,GKTT,BB,DDFK}.
Concrete examples of moduli stabilization 
which work outside the framework of low-energy
supersymmetry have also been developed \cite{EvaAlex}.
In the present paper, we describe a class of type IIA flux vacua
based on Calabi-Yau compactification, where the fluxes alone generate
a potential for all geometrical moduli.

One of the basic difficulties with analyzing ${\cal N}=1$ string
compactifications is that the scalar potential receives corrections
at all orders in $\alpha^\prime$ and $g_s$.
Since one is usually ignorant of the exact K\"ahler potential $K$,
i.e. the exact two-derivative Lagrangian, corrections to the
potential which are suppressed by the ratio of the SUSY breaking
scale $M^2$ to $M_{s}^2$
and which are difficult to compute will typically arise.  For instance,
in a background where a chiral field $\phi$ has a SUSY-breaking
auxiliary field VEV $F_{\phi} \neq 0$, one can potentially soak
up the superspace $\theta$ integrals in $\int d^4\theta \,K$ by
using $F_{\phi}$,
and hence terms in $K$ proportional to $\phi^\dagger \phi$
can correct the scalar potential $V$.
This is not necessarily a problem, since in models where one obtains
moderately weak couplings by tuning, such
corrections can be controlled (as in standard perturbative quantum
field theory) -- this is the situation for many of
the proposed string constructions with stabilized moduli.
Nevertheless, this does represent a concrete limitation on one's knowledge
of the potential.

In the class of models we describe here, in contrast,
one can hope to compute the
exact two-derivative Lagrangian (though it is still a highly
nontrivial task).  These models are based
on ${\cal N}=2$ gauged supergravities; while such
models cannot be completely realistic,
they can be useful toy models for the more general
${\cal N}=1$ situation. Among the many developments during the duality revolution of the
mid 90s
was the discovery that ${\cal N}=2$ supersymmetric string vacua
are in some sense exactly soluble \cite{Ntwoexact}.
Using heterotic/type II duality as well as mirror symmetry,
one can find the exact prepotential for the ${\cal N}=2$ vector
multiplets at string tree-level in the type II picture,
and the geometry of the quaternionic manifold
at string tree-level in the heterotic picture.
In practice this can be carried out for the vector multiplets in
simple examples (see e.g. \cite{KKLMV} where the Seiberg-Witten
solution of ${\cal N}=2$ gauge theory \cite{SW}
was recovered using these dualities).  On the other hand, it
has proved
dauntingly difficult to understand the geometry of hypermultiplet
moduli spaces (see e.g. \cite{Paul} for some attempts).
This is partially because in the IIB picture the
vector multiplet moduli space is exact at both string and
sigma model tree-level, while even in the heterotic picture the
quaternionic manifold receives corrections in sigma model
perturbation theory (though not from string loops).
Nevertheless, $\N=2$ vacua are clearly under
better control than their ${\cal N}=1$
counterparts.

In this paper, we describe a class of type IIA Calabi-Yau compactifications
where a potential for all geometrical moduli (as well as the axio-dilaton)
is determined completely by the ${\cal N}=2$ prepotential and quaternionic
metric, as well as a choice of gauging.  In string theory terms, these
models arise from IIA compactifications on Calabi-Yau spaces with
the RR four-form flux $F_4$ and the NS three-form flux $H_3$ turned on.
While the relevance of gauged supergravity to flux potentials in
string theory has been discussed extensively
\cite{PS,Michelson,louismicu,gauging,Kounnas}, most of the compact
Calabi-Yau flux models which have been constructed
to date also include orientifold planes.
The presence of the orientifold planes
breaks the supersymmetry to ${\cal N}=1$, and so the results derived
from the gauged supergravity analysis are not exact (though they
can be an excellent approximation).  In particular, the
K\"ahler potential need not be the one which follows from
gauged supergravity.  In our class of models, the
results from gauging (in terms of the fully corrected prepotential
and quaternionic metric) should be an even better approximation to
the full theory.  Since in the ${\cal N}=2$ models we
know the precise K\"ahler potential,
one expects the first corrections to the potential that we are
neglecting to be down by two additional powers of $M_s$, in
comparison with typical ${\cal N}=1$ models.

Our models are simpler, more
computable relatives of
$G_2$ flux compactifications of M-theory.  Models with only four-form flux
turned on in that context do not yield
vacua in the large volume approximation, a fact we will see
reflected here as well; orientifolds of our IIA models should provide  
insight into quantum corrections which generate more structure in  
the effective potential in such compactifications. 
Earlier
suggestions for moduli stabilization
in M-theory compactifications appeared in \cite{Acharya}.
Some interesting Calabi-Yau compactifications of IIA strings with only
RR flux turned on were considered in \cite{PS}, while
a different class of
type IIA compactifications with flux was described in \cite{behrndt}.

The organization of this paper is as follows.  In \S2, we describe
the general gauged supergravity framework which encompasses our
models.  In \S3, we present a toy example  
which shows that the resulting potentials can have interesting features.
We close with a discussion in \S4. Our calculations have been
relegated to several appendices.

\newpage
 
\section{Scalar Potentials from Gauging} \label{gauging}

In this section, we describe the scalar potentials which arise
in IIA Calabi-Yau compactifications with RR four-form flux (and six-form flux with qualifications, see below)
and NS three-form flux turned on in the internal dimensions.
In \S\ref{gaugedsugra}, we describe how the potentials are derived given the
data of a 4d ${\cal N}=2$ supersymmetric effective field theory. In \S\ref{correctionstotree} and \S\ref{reduction}, we describe how IIA string compactifications on Calabi-Yau
threefolds with nontrivial $F_4$ and $H_3$ give examples
of the class of theories described in \S\ref{gaugedsugra}.
The latter analysis relies heavily on earlier work of Louis
and Micu \cite{louismicu}.

\subsection{Scalar potentials from ${\cal N}=2$ data} \label{gaugedsugra}

The data of an ${\cal N}=2$ supergravity theory in four dimensions
includes a special K\"ahler manifold ${\cal M}_V$, the moduli space
of vector multiplets, and a quaternionic K\"ahler manifold ${\cal M}_H$,
the moduli space of hypermultiplets.
In type IIA compactification on a Calabi-Yau space $X$, these correspond
to the K\"ahler moduli space and the complex structure + axio-dilaton
moduli space respectively
(with the latter enhanced to a quaternionic manifold by the presence
of the RR axions).

The geometry of ${\cal M}_V$ can be described by
complex projective coordinates $X^I$, $I=0,\cdots,h^{1,1}(X)$,
and a prepotential $F$.  The K\"ahler
potential on ${\cal M}_V$ is
\be
K ~=~-{\rm log}[i(\overline X^I \F_I - X^I \overline \F_I)]~.
\ee
${\cal M}_H$ comes equipped with a quaternionic metric
$h_{uv}$, where $u,v = 1,\cdots,4(h^{2,1}(X)+1)$.

In an ${\cal N}=2$ gauged supergravity \cite{painful},
we in addition choose Killing vectors $k_I^i$ and
$k_I^u$ which generate the action of the $I$th gauge
field on the vector and hypermultiplet moduli ($I=0$ corresponds
to graviphoton charges; we also note that for abelian gauging, the case we will be considering in this paper, $k_I^i$ of course vanish).
These are related to an SU(2) triplet of Killing prepotentials
${\cal P}_I^x$ (for $x=1,2,3$).
The scalar potential is given in terms of the Killing vectors
and the Killing prepotentials by
\be
\label{scalpot}
V = e^K X^I \bar{X}^J ( g_{i \bar{j}} k_I^i k_J^{\bar{j}} + 4 h_{uv} k_I^u k_J^v) - (\frac{1}{2} (\im  \N)^{-1\,IJ} + 4e^K X^I \bar{X}^J) \calP_I^x \calP_J^x  \,.
\ee
$\im \N$ is the gauge coupling matrix,
\ba
{\cal L}_{kin}^{vec} &=& i (\bar{\N}_{I J} F^{-I} \wedge *F^{-J} - \N_{I J} F^{+I} \wedge *F^{+J}) \\
&=& (\im \N)_{I J} F^{I}\wedge *F^{J} - i (\re \N)_{I J} F^{I}\wedge F^{J} \,.
\ea

The important point for us is the following: the scalar potential
of the resulting theory is completely determined in terms of a choice
of isometries and the data characterizing ${\cal M}_V$ and ${\cal M}_H$.
Hence, although it is difficult work to compute the low-energy
effective action of the
${\cal N}=2$ theory resulting from Calabi-Yau compactification, this
action together with various choices of charges (fluxes) completely
determines the potential (\ref{scalpot}) in this class of models.\footnote{
It could be that at the level of $e^{-1/g_s}$ corrections, the relevant
metric on ${\cal M}_H$ is changed by the presence of fluxes.  This is
particularly plausible because Euclidean D2 branes which give rise
to such corrections, cannot wrap cycles threaded by NS three-form flux.
The expected size of the $e^{-1/g_s}$ corrections 
can be made negligibly small by tuning $g_s$ as we
describe explicitly in \S3.1, and hence these effects play no
significant role in our considerations.}

Most
Calabi-Yau flux compactifications to date have involved, in addition to fluxes,
further explicit breaking of the supersymmetry (or have involved
highly simplified models, like toroidal or $K3$ orientifolds).
In \S\ref{reduction}, we describe
a class of IIA Calabi-Yau compactifications
where the theory
is really an ${\cal N}=2$ gauged supergravity, and the formula
(\ref{scalpot}) is the full result for the potential.

\subsection{Corrections to the tree level CY potential}\label{correctionstotree}
It is a celebrated result in (ungauged) $\N=2$ supergravity that up to the two derivative level, no
interaction terms can be introduced that involve both
vector multiplets and (neutral) hypermultiplets.
For CY string compactifications,
this implies that
the metric on the vector multiplet scalar manifold receives no string loop corrections, and the metric on whichever scalar manifold coincides with the complex structure moduli space of the CY at tree level (hyper for IIA, vector for IIB) receives no $\alpha'$ corrections. In gauged supergravity, hypers can acquire charges under vectors. The two sectors of the theory are then of course no longer decoupled. Nevertheless, the above conclusions can still be drawn: the structure of the Lagrangian is completely encoded in terms of the same type of $\N=2$ data as before gauging, i.e. the metric on a special K\"ahler manifold for the vectors and on a quaternionic manifold for the hypers (in addition, as pointed out above, killing vectors encoding the isometries of the scalar manifolds that are to be gauged must be specified) \cite{painful}.
As before, the
presence of a potential notwithstanding, no mixing of the
metrics is allowed, and the
same exactness conclusions as above can then be drawn.

In the presence of the fluxes, backreaction corrects the Calabi-Yau geometry
and it is no longer a solution of the equations of motion.
However, there is strong evidence that the result of turning on fluxes
in the Calabi-Yau must, at the level of 4d effective field theory, simply
be to gauge the ${\cal N}=2$ supergravity arising from the model with no flux.  We assume this
to be true (up to the possible caveat about $e^{-1/g_s}$ 
corrections to the metric on ${\cal M}_H$ mentioned above).  Then,
the powerful results of
\cite{painful} allow us to write down the 4d effective field theory resulting from
the corrected geometry in terms of the ${\cal N}=2$ data of
the fluxless model .  
Therefore, even without detailed knowledge
of the structure of the corrected supergravity  or string solution,
we are able to discuss the effective potential of the resulting low-energy
theory in 4d with confidence.

It is important to keep in mind that the ``exact" results from gauged 
${\cal N}=2$
supergravity receive corrections when one expands the resulting
potentials around critical points which break supersymmetry.  In any
systematic attempt to make controlled examples, the relevant parameter
controlling corrections will be $\mu = M/M_s$ where $M$ is the supersymmetry
breaking scale.  Since the first unknown corrections will enter at
higher orders in the $\mu$ expansion for ${\cal N}=2$ gauged models than
for general ${\cal N}=1$ models, we expect these 
models to be a good laboratory
for studying the space of vacua.

\subsection{IIA strings with $F_4$ and $H_3$ flux} \label{reduction}
The class of models we will consider differ from the hitherto popular IIB models in two main regards.
Unlike the situation in IIB, we demonstrate below that tadpole constraints do not force us to break $\N=2$ SUSY explicitly, e.g. by orientifolding, once we turn on fluxes in IIA. This is why we can extend the power of $\N=2$ beyond the traditional scenario without fluxes. Also, in IIA compactifications on CYs, RR and NS fluxes thread cycles of different dimensions. Since the size of even dimensional cycles is controlled by the K\"ahler data of the geometry, and the size of 
middle dimensional cycles is controlled by the complex structure moduli, turning on both types of fluxes gives rise to non-trivial dependence on both complex structure and K\"ahler moduli in the potential already at the perturbative level in $g_s$.\footnote{The `non-trivial' dependence is to be contrasted with the no-scale potentials in
IIB \cite{GKP},
which also have dependence on both types of moduli, the K\"ahler moduli
however only entering trivially via the $e^K$ prefactor in the potential
of ${\cal N}=1$ supergravity.}

\subsubsection{Tadpole constraints}
In type IIB, turning on both RR and NS 3-form flux $F_3$ and $H_3$ gives rise to a D3 brane tadpole visible in  the CS term
\ba
\int C_4 \wedge H_3 \wedge F_3
\ea
in the SUGRA action. We wish to determine whether such tadpoles can arise in IIA. The fluxes in the game are $F_0, F_2, F_4, F_6$ and $H_3$. Only tadpoles for space filling branes lead to inconsistencies due to violation of Gauss' law in the compact space. Hence, we only need to worry about the gauge potentials $C_5, C_7, C_9$.
\begin{itemize}
\item{A $C_5$ tadpole could only arise from a term $\int C_5 \wedge F_2 \wedge H_3$. Since a CY has no non-trivial 5 cycles, such a term cannot arise in a CY compactification.}
\item{A $C_7$ tadpole can arise from a term $\int C_7 \wedge H_3$. Such a term arises from the kinetic term of $\tilde{F}_2 = dC_1 + m B_2$ in massive IIA, which is IIA with $F_0$ flux turned on,
\ba
\int \tilde{F}_2 \wedge * \tilde{F}_2 &\rightarrow& m \int B_2 \wedge *dC_1 \\
 &=& m \int B_2 \wedge dC_7 \\
 &=& -m \int H_3 \wedge C_7 \,.
\ea}
\item{Finally, there are no 1-forms present to give rise to a $C_9$ tadpole.}
\end{itemize}
The only tadpole for space filling branes that can be generated in this setup hence arises in the presence of both $F_0$ and $H_3$ flux. It is a $C_7$ tadpole, corresponding to a space filling D6 brane wrapping a 3 cycle in the CY.
We will avoid this tadpole by simply leaving the RR $F_0$ flux turned off.

\subsubsection{The Chern-Simons term}
Before proceeding with our analysis, we need to address a subtlety.
Upon turning on fluxes, the relation between field strength and gauge potential, $F=dA$, no longer holds. In particular, various incarnations of the CS terms that are related by integration by parts before turning on fluxes are no longer equivalent. In \cite{louismicu}, this issue is dealt with pragmatically by choosing a form of the action in which the compromised gauge potentials do not appear explicitly in the CS term. This is no longer possible once both RR and NS fluxes are turned on, at least not in 10 dimensions.\footnote{Any fears that turning on RR and NS flux simultaneously may not be consistent should be alleviated, at least for the case of turning on both $F_4$ and $H_3$, by the fact that this situation arises upon compactification of M-theory on a circle in a generic $G$-flux background.}

Requiring that the 10 dimensional action upon compactification fits into the constraining harness of $\N=2$ gauged SUGRA, we arrive at the following proposal for the 10d CS term in a Calabi-Yau background in the presence of fluxes:
\ba \label{csproposal}
S_{CS}=
\int \frac{1}{2} (dB + H_3^{flux})\wedge C_3 \wedge dC_3 - B \wedge F_4^{flux} \wedge dC_3 \,.
\ea
This term clearly reduces to the standard CS term in the absence of fluxes. To derive this proposal from first principles, one could, in the spirit of \cite{Witten:1996md}, introduce a CS term on an 11-manifold ${\cal W}$ that has the physical 10 dimensional space ${\cal M}$ as boundary, $\partial {\cal W}={\cal M}$,
\ba
S_{CS} &=& - \frac{1}{2} \int_{\cal W} H_3 \wedge F_4 \wedge F_4  \,.
\ea
Assuming that ${\cal M}$ is of the form $X \times Z$ where $X$
is a Calabi-Yau manifold and $Z$ represents the noncompact dimensions, and that the cohomology of the Calabi-Yau (together with its ring structure) is preserved in ${\cal W}$,
this prescription
gives rise to the 10d CS term (\ref{csproposal}) proposed above.
To elevate this heuristic sketch into a derivation, one must
show that appropriate manifolds ${\cal W}$ exist, and
that the 10d term is independent of which manifold ${\cal W}$ one chooses.
We take the fact that (\ref{csproposal}) reduces to previously proposed
formulae in suitable limits, together with the fact that it gives rise
to a dimensional reduction which fits into the expected supergravity
framework, to be sufficient evidence for our proposal, and leave
a more rigorous argument (perhaps along the lines suggested above) for
future work.

\subsubsection{Gauged supergravities arising from $F_4$ and $H_3$ flux}
The fluxes we have at our disposal are $F_0, F_2, F_4, F_6$ and $H_3$. To avoid the need to cancel a $D6$ brane tadpole, we set $F_0=0$. As Louis and
Micu \cite{louismicu} have demonstrated (in the absence of $H_3$ flux), turning on $F_2$ and $F_4$ simultaneously gives both electric and magnetic charges to the axion $a$ (the ${\cal N}=1$ SUSY partner of the dilaton) under the gauge fields in the vector multiplet. We bypass such complications by also setting $F_2=0$ (we could just as well consider turning on $F_2$ flux and setting $F_4=0$; this merely swaps electric for magnetic charges for the RR axions). By $F_6=*F_4$, turning on $F_6$ flux is equivalent to modifying the {\it spacetime} part of $F_4$. Hence, the presence of $F_6$ flux can be dealt with {\it after} performing the dimensional reduction. We are thus left with reducing in the presence of $F_4$ and $H_3$ flux. We give the details of this calculation in appendix \ref{compreduction}. Here, we state our results.

We consider compactification on a Calabi-Yau $X$ with the fluxes
\ba
F_4^{flux} &=& e_i \tomega^i \,, \\
H_3^{flux} &=& p^A \alpha_A +q_A \beta^A
\ea
turned on, where $\tomega^i$, $i=1,\ldots,h_{1,1}$, are a dual basis for $H^{1,1}(X)$,\footnote{Meaning, if we take $\omega_i$ to be a basis for $H^{1,1}(X,\IZ)$, then $\int_X \omega_i \wedge \tomega^j = \delta_{i}^j$.} 
and $\alpha_A, \beta^A$, $A=0,\ldots, h_{2,1}$, a symplectic basis for $H^3(X)$. In this flux background, the axion $a$ as well as the RR axions $\xi^A, \txi_A$ become charged under the graviphoton. In addition, $a$ also acquires charges under all vector multiplets. Specifically, the killing vectors are given by
\ba
\label{killare}
k_0^a &=&2 n-2 b^i e_i +p^A \tilde{\xi}_A - q_A \xi^A  \,, \label{graviphotonkilling} \\
k_0^{\xi^A} &=& p^A \,, \nonumber \\
k_0^{\tilde{\xi}_A} &=& q_A \,,  \nonumber \\
k_i^a &=& -2e_i \,, \label{vectorkilling}
\ea
where $n \in {\mathbb Z}$ can be interpreted as $F_6$ flux (see \ref{compreduction}). The $b^i$ are the partners of the (metric) K\"ahler moduli on the complexified K\"ahler cone. As expected, the isometries being gauged by the graviphoton and the gauge fields of the vector multiplets commute pairwise.
Comparing to \cite{louismicu}, we see that the most naive assumption holds true: the isometries being gauged upon turning on both RR and NS flux are simply the sum of those gauged upon turning on the fluxes individually.

The potential we obtain from dimensional reduction also follows, as required for consistency, from the general form of the gauged $\N=2$ potential (\ref{scalpot}) with the above choice of killing vectors. It is given by
\ba
\label{redpot}
V =   \frac{e^{4\phi}}{2\K} (\frac{1}{4}g^{ij} e_i e_j  +(n - b^i e_i+ p^A \tilde{\xi}_A -q_A \xi^A )^2)-\frac{e^{2\phi}}{4\K}( q + p \M)( \im \M)^{-1} ( q + p \bar{\M}) \,,\nonumber \\
\ea
where $\K=\frac{1}{8} e^{-K}$ is the volume of $X$ and $g_{ij} = {1\over
4{\cal K}} \int_{X} \omega_i \wedge *\omega_j$.
Here $\M$ is the matrix defined by
\be
\label{mis}
\M_{AB} = {\overline {\cal G}}_{AB} + 2i {({\rm Im}~{\cal G})_{AC} Z^C
({\rm Im}~{\cal G})_{BD} Z^D \over {Z^C ({\rm Im} {\cal G}_{CD}) Z^D}}~,
\ee
where $Z^A$ and ${\cal G}_A$ are the periods and dual periods in a symplectic
basis for $H^3$, and further subscripts on ${\cal G}$ indicate
differentiation with respect to the relevant complex modulus.

In this section we have simply summarized the results of our computations
because they are somewhat lengthy and involved.  
The interested reader can find the derivation of 
the potential (\ref{redpot}) 
from dimensional reduction in appendix \ref{compreduction}, and 
from the general form of the $\N=2$ gauged supergravity potential (given the  
killing vectors (\ref{killare})) in appendix \ref{compsugra}.

\section{Worldsheet instantons and a toy example}

One class of minima of the potential (\ref{redpot}) lie at infinite K\"ahler parameter. This comes as no surprise: whenever the K\"ahler class dependence is purely that of classical geometry, fluxes will drive the geometry to large K\"ahler class, as this causes them to be diluted and reduces their contribution to the energy.
However, as described in \S\ref{correctionstotree},
we need not restrict our attention to the large volume limit; the data
relevant for computing the scalar potential (in particular, the ${\cal N}=2$
prepotential) is available for all values of the K\"ahler class.  So,
we can simply use the full instanton corrected ${\cal N}=2$ data of
the CY model, to compute the scalar potential
(\ref{scalpot}) including worldsheet instanton corrections.  While this
expression for the potential is not exact, the first corrections
we are neglecting are suppressed
by more powers of $M_s$ than in the analogous ${\cal N}=1$
constructions.  Hence, in any case where the SUSY breaking order
parameters are smaller than $M_s$, corrections may be 
controllable.\footnote{We expect that it will probably be easier to
construct families of such examples in orientifolds of our class of models,
where the negative term in the potential coming from the O-plane tensions
plays a helpful role in stabilizing $g_s$ and volume moduli; for a nice
discussion of this in component form see \cite{Eva}.} 

We leave the construction of such classes of examples for future work, and
here provide only a simple illustrative model that demonstrates that
our potentials can contain interesting structure. 
For our toy example, we will join together the complex structure of a rigid CY with the K\"ahler moduli space of a one K\"ahler parameter CY. Our motivation for considering this fictional geometry is to isolate the main features of this class of compactifications without being swamped by too many computational hurdles. Such a geometry would give rise to one 4 dimensional vector multiplet and one (the universal) hypermultiplet. With the killing vectors determined in the previous section and the associated killing prepotentials determined in appendix \ref{compprepot}, the potential for our toy example takes the form
\ba
V= -e^{4\phi} (\im \N)^{-1 \, IJ} a_I a_J - \frac{1}{2}e^{2\phi} (\tp^2+\tq^2) \left( (\im \N)^{-1 \, 00}  + \frac{3}{4 \K} X^0 \bar{X^0} \right) \,,
\ea
where $a_0 =\frac{1}{\sqrt{2}} n + \tp y - \tq x$ and $a_1=\frac{1}{2\sqrt{2}} e_1$. We have here rotated the RR axions and the fluxes to the more convenient basis $
\sqrt{2} \left(
\begin{matrix}
x\\
y \\
\end{matrix}
\right)
=
\left(
\begin{matrix}
\alpha & \beta \\
\gamma & \delta \\
\end{matrix}
\right)
\left(
\begin{matrix}
\xi\\
\txi \\
\end{matrix}
\right) \;,
\frac{1}{\sqrt{2}} \left(
\begin{matrix}
\tp\\
\tq \\
\end{matrix}
\right)
=
\left(
\begin{matrix}
\alpha & \beta \\
\gamma & \delta \\
\end{matrix}
\right)
\left(
\begin{matrix}
p\\
q \\
\end{matrix}
\right) \;,$
where the transformation matrix is a real symplectic matrix whose entries depend on the period matrix of the complex structure moduli space. The explicit entries are given in appendix \ref{compprepot}.

\subsection{The moduli of the universal hypermultiplet}
The dependence of the potential on the scalar fields of the hypermultiplet is very simple. The potential is quadratic in the string coupling and quadratic in the RR axions, which occur only in the combination $a_0 =\frac{1}{\sqrt{2}} n + \tp y - \tq x$. There is no dependence on the axion $a$.

\paragraph{The dilaton:} highlighting the dilaton dependence, the potential takes the form
\ba
\label{dilpot}
V &=& e^{2\phi} (A_2 e^{2\phi} +A_1) \,,
\ea
where
\ba
A_1 &=& - \frac{1}{2} (\tp^2+\tq^2) \left( (\im \N)^{-1 \, 00}  + \frac{3}{4 \K} X^0 \bar{X^0}\right)  \\
A_2 &=& - (\im \N)^{-1 \,IJ} a_I a_J \,.
\ea
For fixed $A_1$ and $A_2$, there is hence always a stationary point at vanishing string coupling ($g_s = e^{\phi}$), where the potential and all of its derivatives with regard to $\phi$ vanish. For positive $A_1$, this point is a minimum. The other stationary point lies at $e^{2\phi} = -\frac{A_1}{2A_2}$.
Since the dilaton is a real scalar field (and hence the string coupling is
always real and positive), this is only a physically acceptable
solution when the RHS of this equation is positive.

The analysis thus hinges on the signs of $A_1$ and $A_2$. Let us consider these in turn. $\im \N$ is the gauge coupling matrix, and hence negative definite (negative rather than positive definite due to its standard definition in SUGRA). $A_2$ is therefore always positive. For the potential to have a minimum away from vanishing string coupling $e^{\phi}=0$, $A_1$ must thus be negative. 
This imposes the following inequality on the $\N =2$ data: 
\ba \label{dilminineq}
4 (\im \N)^{-1 \,00} \K + 3 X^0 \bar{X^0} > 0  \,.
\ea
Note that the choice of the values of the fluxes did not enter into these considerations.  This is a simplification that would not persist in more general
models, but arises because of our extremely simple choice of vector and
hyper moduli spaces.

In the classical limit, in the gauge $X^0=1$, $(\im \N)^{-1 \, 00} \K = -1$, and the inequality is not satisfied. To move into the interior of moduli space, we need to pick a model. We will do so below. First, let us assume that we can find a model with $A_1<0$ in some region of the K\"ahler moduli space, and press on with the analysis. Plugging in the functional dependence of the string coupling on the remaining parameters of the theory at the minimum, we obtain the following expression for the potential:
\ba
V_{\phi}=-\frac{A_1^2}{4A_2} \,.
\ea
The logic of the notation is that the subscript denotes the field that has been eliminated from the potential.

\paragraph{The RR axions:} $V_{\phi}$ depends on the RR axions $x$ and $y$ in the combination $a_0 =\frac{1}{\sqrt{2}} n + \tp y - \tq x$. Minimizing with regard to $a_0$ yields the potential
\ba
V_{\phi,\xi,\txi} &=& \frac{1}{4} \frac{A_1^2 (\im \N)^{-1 \, 00}}{a_1^2 \det \im \N^{-1}} \\
&=& \frac{1}{256}\frac{ ( 4 (\im \N)^{-1 \, 00} \K + 3 X^0 \bar{X^0})^2(\im \N)^{-1 \, 00}}{\K^2 \det \im \N^{-1}} \frac{(\tp^2 + \tq^2)^2}{ a_1^2} \,.
\ea
Since the eigenvalues of the gauge coupling matrix are, away from singular points, always negative, the determinant is strictly positive.
The sign of $V_{\phi,\xi,\txi}$ hence depends on the
sign of $(\im \N)^{-1 \, 00}$. A simple
argument shows that in models with a single K\"ahler parameter, this
must be negative: at large radius, $(\im \N)^{-1 \, 00}$ is negative. If it is positive at some point in moduli space, it must, by continuity and the fact that singular points on the moduli space are of complex codimension one, vanish at some point. We know that the determinant of the gauge coupling matrix is strictly positive away from singular points. The matrix being symmetric, the off diagonal elements contribute negatively to the determinant. Hence, neither of the diagonal elements can vanish anywhere on moduli space, and they must therefore retain their sign throughout moduli space. We conclude that $V_{\phi,\xi,\txi}$ for our toy example will vanish or be negative. Any minimum will hence be 
anti de Sitter, in this approximation. Furthermore, the 
potential is bounded from
below\footnote{The only singular point
on moduli space is the (mirror) conifold point, and $V \to 0$ there as
$g_s$ is driven to vanish. At any finite value of $g_s$, the potential (naively) tends to positive infinity there.
This is related to the fact that the dilaton carries electric charge,
in contrast to the situation studied in \cite{PS}.
The potential is
manifestly finite at any smooth points.} 
and vanishes in the large radius limit. 
So we can conclude that as long as there is some region in 
K\"ahler moduli space where the potential goes negative (as happens
if $A_1 < 0$), it must
attain a minimum.

Once we minimize with regard to $a_0$, we obtain the following flux dependence for the string coupling
\ba
g_s^2 \sim \frac{\tp^2 + \tq^2}{e_1^2} \,.
\ea
By a judicious choice of fluxes,
one can therefore find vacua at weak string coupling.  This feature
will persist in more general models; in the limit where the RR flux
quantum numbers are larger than the NS flux quantum numbers, the coupling
will always be weak.

\subsection{The K\"ahler moduli}
The dependence on K\"ahler moduli enters the potential via the gauge coupling matrix and $X^0$.
For our toy example, we choose the K\"ahler moduli space as that of
the sextic in $W\IP^4_{2,1,1,1,1}$.  Using mirror symmetry, this is equivalent
to the complex structure moduli space of the orbifold of the hypersurface $W=2x_0^3 + x_1^6 + x_2^6 + x_3^6 + x_4^6$ in $W\IP^4_{2,1,1,1,1}$ by
the group $G = \IZ_3 \times \IZ_6^2$ \cite{GP,Klemm:1992tx}. Following the calculational steps outlined in appendix C, we 
obtain the $\N=2$ data 
necessary to write down the potential in an expansion 
around the Landau-Ginzburg point.

\begin{figure}[h]
\begin{center}
\resizebox{10cm}{!}{\includegraphics{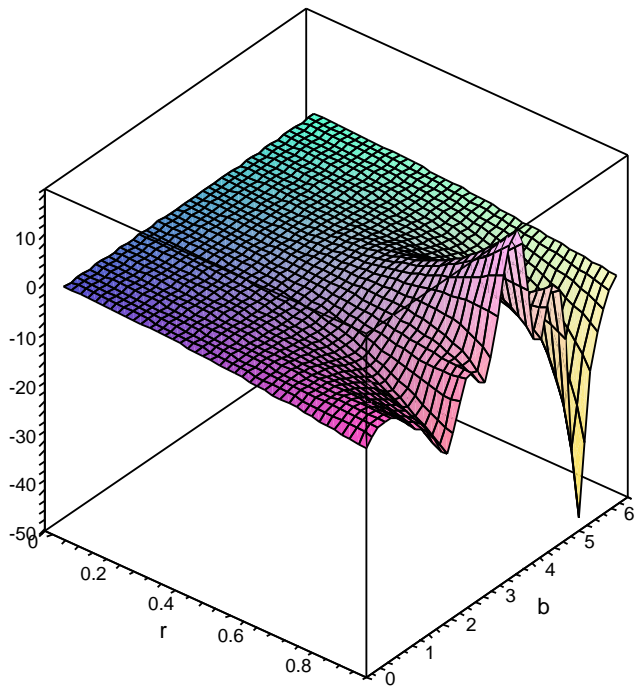}}
\end{center}
\caption{A plot of $4 (\im \N)^{-1 \, 00} \K + 3 X^0 \bar{X^0} $ against $\psi$. To produce this and the following plot, the periods around the LG point were expanded to order 6 in $\psi$.\label{B}}
\end{figure}

As outlined above, the crucial step in our analysis is finding a region in moduli space in which the inequality (\ref{dilminineq}) is satisfied, such that 
$A_1 < 0$ and the minimum of the potential lies away from 
vanishing string coupling.  This is equivalent to requiring that
${\rm (Im}{\cal N}{\rm )}^{-1 00} + {3\over 4{\cal K}} X^0 \bar X^0 > 0$. 
A plot of the LHS of this inequality over the $\psi$ plane, where $\psi = r e^{ib}$ is the coordinate on the K\"ahler moduli space in a neighborhood of the Landau-Ginzburg point (see \ref{period}), is shown in figure \figref{B}.

\begin{figure}[h]
\begin{center}
\resizebox{10cm}{!}{\includegraphics{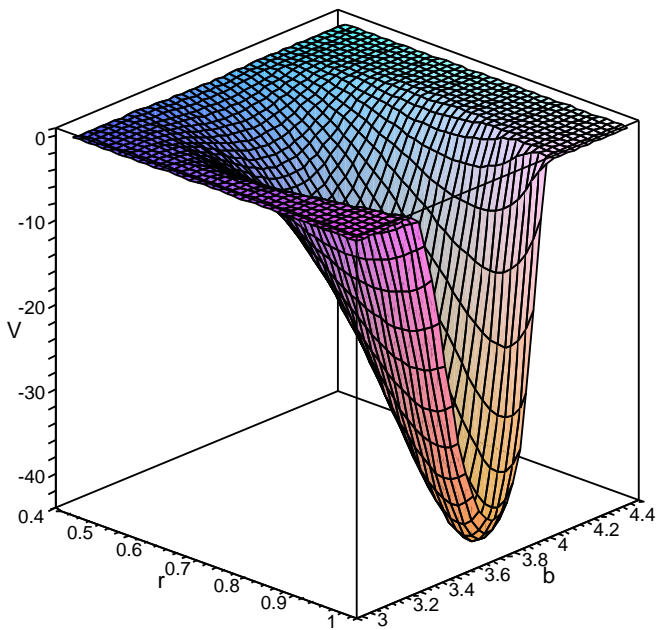}}
\end{center}
\caption{A plot of $V_{\phi,\xi,\txi}$ against $\psi$.\label{V}}
\end{figure}

Computing (\ref{dilpot}) precisely at the Landau-Ginzburg point, we see that 
the minimum of the potential in the dilaton direction would arise by
taking $g_s \to 0$ (where $V$ vanishes).  
On the other hand, as we vary the expectation value of the 
K\"ahler mode by hand, we see that for $r \sim 0.6$ and 
a range of values of the axion, 
the dilaton vacuum arises at a finite value of $g_s$.  
At this critical
point $V < 0$. As 
explained in \S3.1, this is all we need to 
know to infer the existence of a minimum of the full
potential at finite $r$.  To examine whether this minimum lies within our range of computability, we consider a plot of $V_{\phi,\xi,\txi}$ in figure \figref{V}.
The plot suggests that a minimum of the potential lies very close to $r=1$, (but at $b\sim 3.7$, i.e. far from the conifold). This result should be verified by expanding the periods to higher order in $\psi$. Again, we can infer the {\it existence} of a minimum simply by observing the potential take on negative values, and this happens already at $r\sim 0.6$, where we trust our calculation.

Note that in computing the potential, we allow the NS axion partner of
the K\"ahler mode to vary over a full $2\pi$ period; this results in
a smooth potential.  While one might naively expect the axion to vary
only over a pie-wedge of angle ${2\pi \over 6}$ in this model, the
fluxes spontaneously break the $\IZ_6$ symmetry responsible for this
identification.  Hence upon leaving one pie-wedge and entering the
neighboring region, one can either perform a modular transformation
(determined by the LG monodromy) which changes the fluxes (and is
necessary to obtain a smooth potential), or one can fix the fluxes
and allow the axion to vary over a larger region in field space.
We choose the latter course; since the order of the monodromy is 6,
this simply re-enlarges the axion moduli space to have $2\pi$ period.

\subsection{Incorporating string loop corrections}
Since we can dial the size of the string coupling by a diligent choice of fluxes, we can meaningfully incorporate the first string loop corrections to our result. Recall that at the two derivative level, the vector multiplet moduli space receives no corrections at higher loops in $g_s$. The hypermultiplet moduli space receives both perturbative and non-perturbative corrections. The perturbative corrections in the case of a single hypermultiplet have
been studied in e.g. \cite{Antoniadis:1997eg,Andy,Antoniadis:2003sw}.

Our main interest is whether the additional features introduced
by the known loop corrections can qualitatively change the behavior of
the potential in the region of K\"ahler moduli space
where $A_1 > 0$; recall that in the absence
of corrections, vacua at finite $g_s$ only arise if $A_1 < 0$. We will find that the $A_1 >0$ region in parameter space can be redeemed if two other inequalities, which depend sensitively on the K\"ahler {\it and} the flux data, are satisfied.

The potential takes the form
\ba
\label{simpleform}
V = A_1 e^{2\phi} + A_2 e^{4\phi} + A_3 e^{6\phi} \,,
\ea
where
\ba
A_1 &=& - \bigg(  (\im \N)^{-1 \, 00} + \frac{3}{4\K} X^0 \bar{X^0}\bigg) (\tp+\tq)^2\\
A_2 &=& - (\im \N)^{-1 \,IJ} a_I a_J +\chi_1\bigg( (\im \N)^{-1 \, 00} + \frac{1}{2\K}X^0 \bar{X^0} \bigg)(\tp+\tq)^2 \\
A_3 &=&   \frac{\chi_1}{\K}  X^I \bar{X^J}a_I a_J \,,
\ea
and $\chi_1 = \frac{4\zeta(2)\chi}{(2\pi)^3}$, where  $\chi=2$, the Euler number of our make-believe rigid Calabi-Yau. Note that we must distinguish between the terms in the potential stemming from the NSNS and the RR sector of string theory. The latter are kept to one higher order in $e^{2\phi}$.

Let's focus our attention on the string coupling dependence again. The minima of the potential with regard to $g_s$ lie at
\ba \label{minfordil}
\frac{-A_2 \pm \sqrt{A_2^2 - 3 A_1 A_3}}{3 A_3} \,,
\ea
if this is positive, else at $0$.
Again, our analysis boils down to the signs of the three coefficients $A_i$. $A_3$ is positive. The signs of $A_1$ and $A_2$ depend on where we are in the K\"ahler moduli space. First, let's consider the case for which minima away from vanishing string coupling existed at tree level, $A_1 <0$.
A quick glance at (\ref{minfordil}) convinces us that the same is true here, independent of the sign of $A_2$. Next, consider $A_1 >0$. Unlike the situation at tree level, a minimum at finite string coupling is now possible in this region of parameter space, if the inequalities $A_2<0$ and $A_2^2 - 3 A_1 A_3 >0$ can be satisfied. The first contribution to $A_2$ is positive definite. The sign of the second depends on where we are in K\"ahler moduli space, but $A_1>0$ implies that this term is negative. Hence, by choosing the magnitude of the fluxes $\tp,\tq$ carefully compared to that of $n, e_1$, we can arrange for $A_2<0$. Notice that unlike the case at tree level, the fluxes enter crucially in this analysis. Incorporating the second inequality into our considerations requires launching a numeric study of the K\"ahler data similar to our analysis of the tree level inequality above. We leave this to the interested reader.

The lesson we glean from this study is that, as expected,
the computable quantum corrections to $V$ allowed by ${\cal N}=2$
supergravity give rise to interesting
substructure in our potentials.
Note also that, once one has incorporated the positive term $A_3$ into
the potential (\ref{simpleform}), this toy model can potentially 
admit de Sitter vacua,
which was not possible in the approximation of \S3.1.

\section{Discussion}

Our construction provides another illustration of the fact
that fairly common string
theory backgrounds can include enough different effects in the potential
to stabilize the geometric moduli.  Our toy model was
not sufficiently complicated to admit parametric control; 
it would be interesting to
find an analogue of the tuning parameter $W_0$ of \cite{KKLT}, which
would allow one to stabilize some
small fraction of the models in the semi-classical
regime.
Since the unknown corrections to the potential in this class of models
are down by more powers of $M_s$ than in typical ${\cal N}=1$ constructions,
but the potential is nevertheless a reasonably
generic function of all moduli,
one expects this class of models to be a good toy laboratory for
studying moduli stabilization.

In the IIB context, it has recently become clear that the space of
flux vacua admits a statistical description \cite{IIBstat}.
The class of IIA models we study in this paper seems even more amenable to such analysis,
since the potential (\ref{scalpot})
relevant to compactification on a given Calabi-Yau
manifold $X$ can be completely
constructed in terms of classical geometric data on $X$ and its
mirror manifold $Y$.
In other words, this is a setting in
which, in the leading approximation, mirror
symmetry allows one to compute potentials for all moduli.

One should be able to generalize this construction to
a class of ${\cal N}=1$ Calabi-Yau orientifold models where all
geometric moduli enjoy flux-generated potentials.  In that context,
the leading approximation to the potential would still be given
by gauged supergravity formulae (as in the type IIB orientifolds).
We note here that the negative contribution to the scalar potential
coming from the inclusion of orientifold planes in the ${\cal N}=1$
setting should actually make the stabilization of moduli considerably
simpler there than in the ${\cal N}=2$ constructions presented here (for
discussions of the helpful role of orientifold planes, see \cite{GKP}
and \cite{Eva}).
Furthermore, there has recently been significant progress in both
constructing semi-realistic chiral brane models, and in
combining them with flux compactifications (see e.g.
\cite{realistic,softsusy,shiu,cvetic,blumenhagen}).
Many of these constructions arise in IIA string theory, and
should naturally admit embeddings into (orientifolds of) our class
of Calabi-Yau flux models.

Finally, the connection between gauged supergravity and string
compactification fairly
begs the question: which class of string compactifications is generic 
enough to yield the most
general (or even most general abelian) gaugings imaginable?   
We leave these as promising directions for future work.

\begin{center}
{\bf Acknowledgements}
\end{center}
\medskip
We would like to thank S. Ferrara, A. Giryavets, J. Hsu, R. Kallosh, L. McAllister, 
A. Micu, E. Silverstein, T. Taylor, D. Tong, H. Verlinde
and especially K. Behrndt for helpful discussions.
We also thank J. Maldacena for helpful comments about the paper.
The work of S.K. was supported by a David and Lucile Packard
Foundation Fellowship for Science and Engineering, the
Department of Energy, and the National Science
Foundation. The work of AK was supported by the U.S. Department of Energy under contract number DE-AC02-76SF00515.

\appendix

\section{The reduction}
The starting point of our analysis is the following ten-dimensional
type IIA supergravity action,
\ba
S&=& \int e^{-2 \hphi} (\frac{1}{2} R *1 + 2 d\hphi \wedge * d\hphi - \frac{1}{4} H_3 \wedge *H_3) \nonumber \\
& & - \frac{1}{2} (F_2 \wedge * F_2 +  \tilF_4 \wedge * \tilF_4) \nonumber \\
& & + \frac{1}{2} H_3 \wedge C_3 \wedge dC_3 - B \wedge F_4^{flux} \wedge dC_3 \,, \label{action}
\ea
where
\ba
H_3&=& dB + H_3^{flux} \,, \\
F_2 &=& dC_1 \,, \\
F_4 &=& dC_3 + F_4^{flux} \,, \\
\tilF_4 &=& F_4 - C_1 \wedge H_3 \,.
\ea
As reviewed in section \ref{gauging}, the 4d effective theory of the compactified 10d SUGRA becomes gauged upon turning on fluxes. Gauged $\N=2$ SUGRA in 4d exhibits a potential which is completely determined in terms of the $\N=2$ data of the ungauged theory and the killing vectors of the isometries that are being gauged.
Our goal in this appendix is to derive the potential as well as the killing vectors from dimensional reduction, and demonstrate that they fit the expected $\N=2$ mold.

\subsection{Potential and killing vectors from reduction} \label{compreduction}
We consider compactification on a Calabi-Yau $X$. We begin by decomposing the field strengths as follows,
\ba
F_4= dc_3 + dA^i \omega_i + d\xi^A \alpha_A + d\tilde{\xi}_A \beta^A + e_i \tomega^i \, \\
H_3 = db_2 +db^i \omega_i + p^A \alpha_A + q_A \beta^A \,.
\ea
Here, $\omega_i$, $i=1,\ldots,h_{1,1},$ are a basis for $H^{1,1}(X)$, and $\tomega^i$ a dual basis spanning $H^{2,2}(X)$, i.e.
\ba
\int_X \omega_i \wedge \tomega^j = \delta_i^j  \,.
\ea
$\alpha_A, \beta^A$, $A=0,\ldots, h_{2,1},$ are a symplectic basis for $H^3(X)$,
\ba
\int_X \alpha_A \wedge \beta^B = \delta_A^B \,.
\ea

Contributions to the potential arise from the kinetic terms of $C_3$,
\ba
- \frac{1}{2} (\tilde{F}_4 \wedge * \tilde{F}_4) &\rightarrow& -\frac{1}{8\K} e_i e_j  g^{ij}\,,
\ea
and the kinetic term for $B_2$,
\ba
\lefteqn{   -\frac{e^{-2\hphi}}{4}H_3 \wedge * H_3} \\
&\rightarrow& -\frac{e^{-2\hphi}}{4} (p^A p^B \alpha_A \wedge *\alpha_B + p^A q_B (\alpha_A \wedge *\beta^B + \beta^B \wedge *\alpha_A)+ q_A q_B \beta^A \wedge *\beta^B)  \,. \nonumber
\ea
Upon expressing the integrals over the 3 cycles in terms of the period matrix $\M$, and introducing the 4d dilaton $e^{-2\phi} = e^{-2\hphi} \K$, this last expression becomes
\ba
 \frac{e^{-2\phi}}{4 \K} ( q + p \M)( \im  \M)^{-1} ( q + p \bar{\M}) \,.
\ea

In addition, the spacetime field $c_3$ turns out to play an important role in determining the potential. We collect all terms containing $c_3$,
\ba
\lefteqn{- \frac{1}{2} (\tilde{F}_4 \wedge * \tilde{F}_4) + \frac{1}{2} H_3 \wedge C_3 \wedge dC_3  - B \wedge F_4^{flux} \wedge dC_3} \\
&\rightarrow& -\frac{\K}{2}(dc_3 - A \wedge db_2) \wedge *(dc_3 - A \wedge db_2) +\\
& &( p^A \tilde{\xi}_A - q_A \xi^A - b^i e_i)dc_3  \,.  \label{dc3}
\ea
$A$ here is the 4d graviphoton field which descends from $C_1$. 
$dc_3$, being dual to a 0 form field strength, carries no local degrees of freedom. We can eliminate it from the action, following \cite{beasleywitten, louismicu}, by solving for it via its equations of motion and plugging back into the action,
\ba
-\int P (dc_3 -j)\wedge * (dc_3 -j)+ Q \,dc_3 \rightarrow -\int \frac{1}{4P}(Q+n)^2 + (Q+n) j \,.
\ea
Here, $n$ is an integration constant which must be chosen to be integral
\cite{boussopolchinski}. This choice of integer in the potential is the exact analogue of the integer appearing in the potential of the massive Schwinger model studied by Coleman \cite{Coleman, beasleywitten}. The part of codimension 2 objects charged under the top form which are nucleated out of the vacuum is here played by $D2$ branes in spacetime. Since $F_6$ flux jumps in between a $D2$/
${\overline{D2}}$ pair, we see that the choice of
$n$ from a 10d point of view corresponds to the choice of $F_6$ flux.

Returning to our task, we apply the above to (\ref{dc3}),
\ba
-\frac{1}{2\K}(- b^i e_i +p^A \tilde{\xi}_A - q_A \xi^A+n)^2 +(- b^i e_i +p^A \tilde{\xi}_A - q_A \xi^A+n) A \wedge db_2 \,.
\ea
Collecting terms and passing to the Einstein frame, $\sqrt{g} \rightarrow \sqrt{g} e^{4\phi}$, we arrive at the 4d potential
\ba
V =   \frac{e^{4\phi}}{2\K} (\frac{1}{4}g^{ij} e_i e_j  +(n - b^i e_i+ p^A \tilde{\xi}_A -q_A \xi^A )^2)-\frac{e^{2\phi}}{4\K}( q + p \M)( \im \M)^{-1} ( q + p \bar{\M}) \,. \nonumber \\ \label{aredpot}
\ea

Next, we turn towards determining the killing vectors of the isometries being gauged. These can be read off from the covariantized kinetic terms of the 4d fields which acquire charges under gauging. To obtain the kinetic term for the axion $da = *db_2$, let us collect all terms involving $b_2$,
\ba
\lefteqn{-\frac{e^{-2\phi}\K}{4}db_2 \wedge *db_2 +(- b^i e_i +p^A \tilde{\xi}_A - q_A \xi^A+n) A \wedge db_2 + \frac{1}{2}db_2 \wedge (\tilde{\xi}_A d\xi^A-\xi^A d\tilde{\xi}_A  + 2e_i A^i)}  \nonumber\\
&=& -\frac{e^{-2\phi}\K}{4}db_2 \wedge *db_2 +\nonumber \\
& &\frac{1}{2}db_2 \wedge [(2b^i e_i -2n-p^A \tilde{\xi}_A +q_A \xi^A) A +2e_i A^i + \tilde{\xi}_A (d\xi^A - p^AA) - \xi^A (d\tilde{\xi}_A -q_AA)] \,. \nonumber
\ea
Dualizing $b_2$ along the lines of
\ba
-\int [P(db_2 \wedge *db_2) - \frac{1}{2}db_2 \wedge j ] \rightarrow -\int \frac{1}{16P} (da +j) \wedge * (da +j)
\ea
yields
\ba
 -\frac{e^{2\phi}}{4\K} [da +(2b^i e_i -2n-p^A \tilde{\xi}_A + q_A \xi^A) A +2e_i A^i + \tilde{\xi}_A (d\xi^A - p^AA) - \xi^A (d\tilde{\xi}_A -q_AA)]^2 \nonumber\\
 =-\frac{e^{2\phi}}{4\K} [Da + \tilde{\xi}_A D\xi^A - \xi^A D\tilde{\xi}_A] ^2 \,\,\,\,\,\,\,\,\,
 \ea
with
\ba
Da &=& da +(2b^i e_i -2n-p^A \tilde{\xi}_A + q_A \xi^A) A +2e_i A^i \,, \\
D\xi^A &=& d\xi^A - p^A A\,,\\
D\tilde{\xi}_A &=& d\tilde{\xi}_A -q_A A\,.
\ea
Hence, we see that, as promised in section \ref{gauging}, turning on $H_3$ and $F_4$ leads to the gauging of the following isometries,
\ba  \label{killing}
k_0^a &=&2 n-2 b^i e_i +p^A \tilde{\xi}_A - q_A \xi^A  \,, \\
k_0^{\xi^A} &=& p^A \,, \nonumber\\
k_0^{\tilde{\xi}_A} &=& q_A \,, \nonumber \\
k_i^a &=& -2e_i \,.
\ea

\subsection{Consistency with gauged SUGRA} \label{compsugra}
The most general scalar potential in $\N=2$ gauged supergravity with only electric charges is given by
\ba \label{generalpot}
V = e^K X^I \bar{X}^J ( g_{i \bar{j}} k_I^i k_J^{\bar{j}} + 4 h_{uv} k_I^u k_J^v) - (\frac{1}{2} (\im  \N)^{-1\,IJ} + 4e^K X^I \bar{X}^J) \calP_I^x \calP_J^x  \,.
\ea
Let's take a closer look at the various ingredients in turn. $g_{i\bar{j}}$, $i,\bar{j}=1,\ldots, n_V$, and $h_{uv}$, $u,v=1,\ldots, 4n_H$, are the metrics on the vector and hypermultiplet moduli space, $\M_V$ and $\M_H$, respectively. The tree level metric on $\M_H$ was derived in \cite{Cecotti:1988qn} from dimensional reduction to be
\ba
ds^2 &=& d \phi \otimes d \phi + \frac{e^{4\phi}}{4}\left[da + \txi_A d\xi^A - \xi^A d\txi_A\right]\otimes \left[da + \txi_A d\xi^A - \xi^A d\txi_A\right]  \nonumber \\
& &-\frac{e^{2\phi}}{2} (\im \M^{-1})^{AB}\left[d\txi_A  +\M_{AC}d\xi^C \right] \otimes   \left[d\txi_B  +\overline{\M}_{BD}d\xi^D \right] \,. \nonumber
\ea
$k^i_I$, $I=0,\ldots, n_V$, are the components of killing vectors encoding isometries of $\mv$, gauged by the gauge field from the $I$th vector multiplet, $I=0$ denoting the graviphoton of the gravity multiplet. Due to $\N=2$ supersymmetry, gauging isometries of $\mv$ implies introducing non-abelian gauge symmetries. These isometries hence cannot become gauged merely by turning on fluxes. The $k^u_I$ are the killing vectors on $\mh$ we determined above, and the $\calP_I^x$ are the corresponding prepotentials, which we shall compute for the case of $n_H=1$ in the next subsection. For the analysis immediately below, we will only need to know $\calP_i^x = -e^{2\phi} e_i \delta^{x3}$.

$\im {\cal N}$ is the gauge coupling matrix and $\re {\cal N}$ the $\theta$ angle matrix for the vectors and graviphoton. When the metric on the K\"ahler moduli space is encoded in a prepotential $\F$, $\N$ is related to $\F$ by
\ba
\N_{IJ} &=& \bar{\F}_{IJ}+ 2i\frac{(\im \F)_{IK}X^K (\im \F)_{JL}X^L}{X^K (\im \F)_{KL} X^L} \,.
\ea
The standard large radius prepotential $\F = -\frac{1}{3!}\frac{\K_{ijk}X^i X^j X^k}{X^0}$ on the complexified K\"ahler moduli space arises as the vector multiplet prepotential upon dimensional reduction of the SUGRA action, {\em not} as we have written it down in (\ref{action}), but after the field redefinition $C_3 \rightarrow C_3 - A_1 \wedge B$. This new basis for the gauge fields is hence more natural when working with a prepotential. The gauge fields in the action (\ref{action}) are related to those in this basis by $A_i \rightarrow A_i - b_i A$. Under this transformation, the killing vectors (\ref{killing}) transform into
\ba
k_0^a &=&2 n +p^A \tilde{\xi}_A - q_A \xi^A  \,, \\
k_0^{\xi^A} &=& p^A \,, \nonumber\\
k_0^{\tilde{\xi}_A} &=& q_A \,, \nonumber \\
k_i^a &=& -2e_i \,.
\ea
Since we will be working in this new basis from now on, we do not bother with introducing new notation to distinguish the two bases.
The advantage of using the transformed killing vectors is that when moving away from the large radius limit, we can simply use the period matrix of the mirror CY as our gauge coupling matrix, without need of transforming into a different basis.

From the prepotential given above, one easily obtains the following inverse gauge coupling matrix \cite{louismicu},
\ba \label{gaugecouplingfromprepot}
(\im \N)^{-1} &=&
-\frac{1}{\K}\left(
\begin{matrix}
1 & b^i \\
b^j & \frac{1}{4}g^{ij}+b^i b^j \\
\end{matrix}
\right) \,,
\ea
where $e^{-K}= 8 \K$.

With these preparations, we now demonstrate that the potential (\ref{aredpot}) is consistent with the general form of the potential (\ref{generalpot}) obtained from gauged supergravity. For readability, we divide the potential into three pieces and evaluate them separately.

\paragraph{Terms involving only the graviphoton killing vectors}
\ba
\frac{1}{2\K}  h_{uv} k_0^u k_0^v&=&\frac{1}{2\K} [\frac{e^{4\phi}}{4}(k^a_0 +\txi_A k^{\xi^A}_0 - \xi^A k^{\txi_A}_0)^2 - \frac{e^{2\phi}}{2}( q + p \M)( \im \, \M)^{-1} ( q + p \bar{\M}) ] \nonumber\\
&=& \frac{1}{2\K} [\frac{e^{4\phi}}{4}(2n +p^A \tilde{\xi}_A - q_A \xi^A +\txi_A p^A - \xi^A q_A)^2 \nonumber\\
& &- \frac{e^{2\phi}}{2}( q + p \M)( \im \, \M)^{-1} ( q + p \bar{\M}) ] \nonumber \\
&=& \frac{1}{2\K} [e^{4\phi}(n+p^A \tilde{\xi}_A - q_A \xi^A )^2 - \frac{e^{2\phi}}{2}( q + p \M)( \im \, \M)^{-1} ( q + p \bar{\M}) ] \,. \nonumber \\ \label{gravi}
\ea

\paragraph{Terms involving only the vector multiplet killing vectors}
\ba
4e^K X^i \bar{X}^j ( h_{uv} k_i^u k_j^v -P_i^x P_j^x)- \frac{1}{2} (\im \, \N)^{-1\,ij} P_i^x P_j^x &=&- \frac{1}{2} (\im \, \N)^{-1\,ij} P_i^x P_j^x \nonumber\\
&=&\frac{e^{4\phi}}{2\K}(\frac{1}{4}g^{ij}+b^i b^j) e_i e_j  \,. \nonumber\\ \label{vector}
\ea

\paragraph{Terms mixing graviphoton and vector multiplet killing vectors}
\ba
e^K (X^i \bar{X}^0 +X^0 \bar{X}^i )4 h_{uv} k_0^u k_i^v  &-& [(\im \, \N)^{-1\,i0} + 4e^K (X^i \bar{X}^0+X^0 \bar{X}^i )] P_i^x P_0^x \nonumber\\
&=& e^K (2b^i)4 h_{uv} k_0^u k_i^v  - [(\im \, \N)^{-1\,i0} + 4e^K (2b^i )] P_i^x P_0^x \nonumber\\
&=& \frac{1}{\K} b^i h_{uv} k_0^u k_i^v  \nonumber \\
&=& -\frac{e^{4\phi}}{\K} (n + p^A \txi_A -q_A \xi^A) b^i e_i  \,, \label{gravivector} \,.
\ea
Adding the three contributions (\ref{gravi}), (\ref{vector}), and (\ref{gravivector}), we arrive, as promised, at the potential (\ref{aredpot}) obtained by dimensional reduction.

\section{The killing prepotential for the universal hypermultiplet} \label{compprepot}
The hypermultiplet moduli space ${\cal M}_H$ is a quaternionic manifold of dimension $4n_H$. The structure group of the tangent bundle is hence the product $Sp(2) \times Sp(2n_H)$, and the Levi-Civita connection and curvature decompose accordingly. Relevant for defining the killing prepotential are the $Sp(2)\cong SU(2)$ connection, $\omega$, and its curvature, $\Omega$. In terms of these, the relation between a killing vector $k = k^a \partial_{q^a}$ encoding an isometry of ${\cal M}_H$ and the corresponding killing prepotential $\calP = \calP^x \sigma_x$ is
\ba \label{killprepot}
\iota_k \Omega^x = D \calP^x + \epsilon^{xyz} \omega^y \calP^z\,.
\ea
Given the metric on ${\cal M}_H$ and a killing vector $k$, one must hence calculate the connection and curvature, extract the $Sp(2)$ factor and then solve the above differential equation to determine the corresponding killing prepotential $\calP$.

For simplicity, we will restrict ourselves to the universal hypermultiplet.
The holonomy can be made explicit with the choice of vierbein
\ba
q=
\left(
\begin{matrix}
u &v\\
\bar{v} & -\bar{u}\\
\end{matrix}
\right) \,,
\ea
in terms of which the metric is given by
\ba \label{univhypmetric}
ds^2 = q^{A\alpha} (\sigma_2)_{AB} (\sigma_2)_{\alpha \beta} q^{B \beta} \,.
\ea
The 1-forms $u$ and $v$ get the following contributions at tree and 1-loop level \cite{Antoniadis:2003sw}
\ba
u_0 &=& e^{\phi} (dx+idy) \\
v_0 &=& -d\phi + e^{2\phi}i(ydx -x dy + \frac{1}{2}da) \\
u_1 &=& - \chi_1 e^{2 \phi} u_0 \\
v_1 &=& - \frac{1}{2} \chi_1 e^{2\phi} v_0 \,,
\ea
where $\chi_1 = \frac{4 \zeta(2) \chi}{(2\pi)^3}$, $\chi$ being the Euler number of the compactification manifold.
The (first) $Sp(2)$ factor of the curvature is given by \cite{Antoniadis:2003sw}

\ba
\Omega^1_0 &=& i(\bar{u_0} \wedge v_0 + \bar{v_0} \wedge u_0) \\
\Omega^2_0 &=& (\bar{u_0} \wedge v_0 - \bar{v_0} \wedge u_0) \\
\Omega^3_0 &=& i(\bar{u_0} \wedge u_0 - \bar{v_0} \wedge v_0)  \\
\Omega^1_1 &=& e^{2\phi} \left( -\chi_1 \Omega^1_0 +i \frac{\chi_1}{2} (u_0 \wedge v_0 - \bar{u_0} \wedge \bar{v_0}) \right) \\
\Omega^2_1 &=& e^{2\phi} \left( -\chi_1 \Omega^2_0 - \frac{\chi_1}{2} (u_0 \wedge v_0 + \bar{u_0} \wedge \bar{v_0}) \right) \\
\Omega^3_1 &=& e^{2\phi} \left( -2 \chi_1 \Omega^3_0 - 2i \chi_1 (\bar{v_0} \wedge v_0) \right)
\ea

The three killing vectors of this metric that correspond to shift symmetries of the axions are
\ba
\mathbf{k_1} &=& \partial_a \\
\mathbf{k_2} &=& 2y \partial_a + \partial_x \\
\mathbf{k_3} &=& -2x \partial_a + \partial_y  \,.
\ea
Note that these shift symmetries are preserved by the perturbative loop corrections to the metric (in fact, this is a crucial ingredient in deriving these corrections).
To calculate the killing prepotential, we use the relation
\ba \label{prepottok}
\calP^x = \frac{1}{4} D^{i}k^j \Omega^x_{ij} \,.
\ea
This yields
\ba
\calP_1 &=&
\frac{1}{2}\left(
\begin{matrix}
0 \\
0 \\
e^{2\phi}-4 \chi_1^2e^{6\phi} \\
\end{matrix}
\right) \\
\calP_2 &=&
2\left(
\begin{matrix}
0\\
- e^{\phi} +\frac{1}{2} \chi_1 e^{3\phi}\\
e^{2\phi} y -4 \chi_1^2 y e^{6\phi}\\
\end{matrix}
\right) \\
\calP_3 &=&
-2\left(
\begin{matrix}
e^{\phi} - \frac{1}{2} \chi_1 e^{3\phi}\\
0\\
 e^{2\phi}x +4 \chi_1^2 x e^{6\phi} \\
\end{matrix}
\right) \,.
\ea

To relate these results to our compactification, we need to identify the coordinates in which we obtain the quaternionic manifold via dimensional reduction to those introduced here. This is easily accomplished by comparing the metric (\ref{univhypmetric}) at tree level to the one obtained via reduction. The latter is
\ba
ds_{red}^2 &=& d \phi^2 + \frac{e^{4\phi}}{4}[da + \txi d\xi - \xi d\txi]^2 -
\frac{e^{2\phi}}{2\M_I}[d\txi^2+\M_R (d\xi d\txi + d\txi d\xi) + |\M|^2 d\xi^2] \,.\nonumber  \\ \label{metis}
\ea
Recall that $\M$ is the period matrix on the complex structure moduli space,
and in (\ref{metis}) we use the 
notation that ${\cal M}_{R,I}$ are its real and imaginary
parts. Since we are considering a rigid Calabi-Yau, $\M$ is simply a constant in our example which encodes the expansion coefficients of the Hodge duals $*\alpha$ and $*\beta$ of a symplectic basis $\{\alpha, \beta\}$ of $H^3$ of the rigid CY in this basis. In particular,
\ba
(\im \M)^{-1} = -\int \beta \wedge * \beta  \,,
\ea
i.e $\im \M < 0$.
Plugging in the tree level expressions for $u$ and $v$ into  (\ref{univhypmetric}) yields
\ba
ds^2 = d\phi^2 + e^{4\phi}{4}(da + 2(ydx-xdy))^2 + e^{2\phi}(dx^2 + dy^2) \,.
\ea
We can read off that $\sqrt{2} \left(
\begin{matrix}
x\\
y \\
\end{matrix}\right)$ and $\left(
\begin{matrix}
\xi\\
\txi \\
\end{matrix}
\right)$ are related via a symplectic transformation. With the parametrization
\ba
\sqrt{2} \left(
\begin{matrix}
x\\
y \\
\end{matrix}
\right)
=
\left(
\begin{matrix}
\alpha & \beta \\
\gamma & \delta \\
\end{matrix}
\right)
\left(
\begin{matrix}
\xi\\
\txi \\
\end{matrix}
\right) \;,
\ea
we determine this transformation to be
\ba
\alpha&=& -\frac{|\M|}{\sqrt{-\M_I}} \\
\beta&=& -\frac{M_R}{|\M|\sqrt{-\M_I}} \\
\gamma &=& 0 \\
\delta&=&\frac{\sqrt{-\M_I}}{|\M|} \,.
\ea

This field identification remains correct at 1-loop level.
By the linearity of (\ref{prepottok}), expanding the isometries
 gauged by the graviphoton (\ref{graviphotonkilling}) and the vectors in the vector multiplets (\ref{vectorkilling}) in this set of killing vectors allows us to read off the corresponding killing prepotentials. The expansions are easily determined to be
\ba
\mathbf{k_{grav}} &=& (2n+ p \txi - q \xi) \partial_a + p \partial_{\xi} + q \partial_{\txi} \\
&=& 2n \mathbf{k_1} + \frac{1}{\sqrt{2}}(\alpha p+ \beta q) \mathbf{k_2} +  \frac{1}{\sqrt{2}}\delta q  \mathbf{k_3} \,,  \\
\mathbf{k_{vect}} &=& -2 e \mathbf{k_1} \,.
\ea

\section{Calculations at the Landau-Ginzburg point}
For our example, we choose the CY given as a hypersurface in the weighted projective space $W\IP^4_{2,1,1,1,1}$. The periods around the Landau-Ginzburg (LG) 
point can be determined in terms of the power series
\ba \label{period}
\omega_0(\psi) &=& -\frac{1}{k \pi^4} \sum_{n=1}^{\infty} \frac{\prod_{i=0}^{4} \Gamma(\frac{n}{k} \nu_i) \sin(\frac{\pi n}{k} \nu_i)}{\Gamma(n)}\frac{e^{i \frac{\pi}{k}(k-1)n}}{\sin(\frac{\pi n}{k})} (\gamma \psi)^n  \,,
\ea
where the $\nu_i$ are the weights of the ambient weighted projective space, $k=6$ (this is the smallest common multiple of the powers $n_i$ in the polynomial $\sum_{i=0}^{4} x_i^{n_i}$ defining the hypersurface; e.g. $k=5$ for the quintic), and $\gamma= k \prod_{i=0}^{4} (\nu_i)^{-\nu_i/k}$.
A basis for the solutions to the Picard-Fuchs equations is now given by $\{\omega_0,\omega_1,\omega_2,\omega_5\}$, where $\omega_i(\psi)= \omega_0(\beta^j \psi)$ for $\beta = \exp(\frac{2\pi i}{k})$. Assembling these in a vector $\omega = -\frac{(2\pi)i^3}{Ord G} (\omega_2, \omega_1, \omega_0, \omega_5)^T$, they are related to a symplectic basis $\Pi'=(\G_1,\G_2,z^1,z^2)^T$ via the transformation $\Pi'=m \omega$, with $m$ given by
\ba
m=
\left(
\begin{matrix}
-\frac{1}{3} &-\frac{1}{3}& \frac{1}{3}& \frac{1}{3}\\
0 &0& -1& 0\\
-1 &0 &3 &2\\
0 &1& -1& 0\\
\end{matrix}
\right)  \,.
\ea

To calculate the gauge coupling matrix, we need the second derivatives of the prepotential $\G$. Using the homogeneity of $\G$ these can be expressed in terms of the periods as follows,
\ba
\G_{12} &=& \frac{\G_1' -\frac{\G_1}{z^1} (z^1)'}{(z^2)'-\frac{z^2}{z^1} (z^1)'} \\
\G_{11} &=& \frac{\G_1}{z^1} - \G_{12} \frac{z^2}{z^1} \\
\G_{22} &=& \frac{\G_2}{z^2} - \G_{12} \frac{z^1}{z^2} \,,
\ea
where the prime denotes differentiation with regard to $\psi$.
We next need to relate these results to the large radius limit. With the period vector in the large radius limit given as $\Pi \sim (t^2, t^3, t,1)^T$, our choice of the matrix $N$ relating the two bases (recall that this matrix is only specified up to monodromy) is given by
\ba
N=
\left(
\begin{matrix}
0 &0& -1& 0\\
0 &0& 0& -1\\
1 &0 &0 &0\\
0 &1& 0& 0\\
\end{matrix}
\right)
\ea
After calculating the gauge coupling matrix in the basis $\Pi'$, we can transform it to the basis $\Pi$: given a symplectic transformation $
N=
\left(
\begin{matrix}
A &B\\
C &D\\
\end{matrix}
\right) $,
the gauge coupling matrix $\N$ transforms as $\N'=\big(B+ A\N(\psi)\big) \big(D+C\N(\psi)\big)^{-1}$.

\renewcommand{\Large}{\large}


\begin{thebibliography}{99}

\bibitem{oldhet1}
J. Derendinger, L. Ibanez and H. Nilles, ``On the low energy $D=4$, ${\cal N}=1$
supergravity theory extracted from the $D=10$, ${\cal N}=1$ superstring,''
Phys. Lett. {\bf B155} (1985) 65;\\
M. Dine, R. Rohm, N. Seiberg and E. Witten, ``Gluino condensation in
superstring models,'' Phys. Lett. {\bf B156} (1985) 55.

\bibitem{oldhet2}
N. Krasnikov, ``On supersymmetry breaking in superstring theories,''
Phys. Lett. {\bf B193} (1987) 37;\\
L. Dixon, ``Supersymmetry breaking in string theory,'' in {\it DPF
Conf. 1990: 811};\\
T.R. Taylor, ``Dilaton, gaugino condensation and supersymmetry
breaking,'' Phys. Lett. {\bf B252} (1990) 59.



\bibitem{Eva}
E. Silverstein, ``TASI/PITP/ISS lectures on moduli and
microphysics,'' hep-th/0405068.

\bibitem{reviews} 
V. Balasubramanian, ``Accelerating universes and string theory,''
Class. Quant. Grav. {\bf 21} (2004) S1337, hep-th/0404075;\\
A. Frey, ``Warped strings: Selfdual flux and contemporary
compactifications,'' hep-th/0308156.

\bibitem{IIBflux}
S. Gukov, C. Vafa and E. Witten, ``CFTs from Calabi-Yau fourfolds,''
Nucl. Phys. {\bf B584} (2000) 69, hep-th/9906070;\\
K. Dasgupta, G. Rajesh and S. Sethi, ``M-theory, orientifolds and
G-flux,'' JHEP {\bf 9908} (1999) 023, hep-th/9908088;\\
T.R. Taylor and C. Vafa, ``RR flux on Calabi-Yau and partial
supersymmetry breaking,'' Phys. Lett. {\bf B474} (2000) 130, hep-th/9912152;\\
P. Mayr, ``On supersymmetry breaking in string theory and its realization
in brane worlds,'' Nucl. Phys. {\bf B593} (2001) 99, hep-th/0003198;\\
B. Greene, K. Schalm and G. Shiu, ``Warped compactification in M and
F theory,'' Nucl. Phys. {\bf B584} (2000) 480, hep-th/0004103;\\
G. Curio, A. Klemm, D. L\"ust and S. Theisen, ``Type II String
Compactifications on Calabi-Yau spaces with H-fluxes,''
Nucl. Phys. {\bf B609} (2001) 3, hep-th/0012213;\\
M. Haack and J. Louis, ``M theory compactified on Calabi-Yau fourfolds
with background fluxes,'' Nucl. Phys. {\bf B635} (2002) 395, hep-th/0103068;\\
K. Becker and M. Becker, ``Supersymmetry breaking, M-theory and Fluxes,''
JHEP {\bf 0107} (2001) 038, hep-th/0107044;\\
S. Kachru, M. Schulz and S. Trivedi, ``Moduli stabilization from fluxes
in a simple IIB orientifold,'' JHEP {\bf 0310} (2003) 007, hep-th/0201028;\\
A. Frey and J. Polchinski, ``${\cal N}=3$ warped compactifications,''
Phys. Rev. {\bf D65} (2002) 126009, hep-th/0201029;\\
M. Berg, M. Haack and B. Kors, ``An orientifold with fluxes and
branes via T-duality,'' Nucl. Phys. {\bf B669} (2003) 3, hep-th/0305183;\\
A. Giryavets, S. Kachru, P. Tripathy and S. Trivedi, ``Flux
compactifications on Calabi-Yau threefolds,'' JHEP {\bf 0404} (2004) 003,
hep-th/0312104;\\
A. Saltman and E. Silverstein, ``The scaling of the no-scale
potential and de Sitter model building,'' hep-th/0402135.

\bibitem{GKP}
S. Giddings, S. Kachru and J. Polchinski, ``Hierarchies from fluxes
in string compactifications,''  Phys. Rev. {\bf D66} (2002) 106006,
hep-th/0105097.

\bibitem{KKLT}
S. Kachru, R. Kallosh, A. Linde and S. Trivedi, ``de Sitter vacua
in string theory,'' Phys. Rev. {\bf D68} (2003) 046005, hep-th/0301240.

\bibitem{DDF}
F. Denef, M. Douglas and B. Florea, ``Building a better
racetrack,'' JHEP {\bf 0406} (2004) 034, hep-th/0404257.

\bibitem{GKTT}
L. G\"orlich, S. Kachru, P. Tripathy and S. Trivedi, ``Gaugino
condensation and nonperturbative superpotentials in flux
compactifications,'' hep-th/0407130.

\bibitem{BB}
V. Balasubramanian and P. Berglund,
``Stringy corrections to K\"ahler potentials, SUSY breaking,
and the cosmological constant problem,'' hep-th/0408054.

\bibitem{DDFK}
F. Denef, M. Douglas, B. Florea and S. Kachru, to appear.

\bibitem{EvaAlex}
A. Saltman and E. Silverstein, ``A new handle on de Sitter
compactifications,'' hep-th/0411271;\\
A. Maloney, E. Silverstein and A. Strominger, ``de Sitter space in
noncritical string theory,'' hep-th/0205316.


\bibitem{Ntwoexact}
S. Kachru and C. Vafa, ``Exact Results for ${\cal N}=2$ Compactifications
of Heterotic Strings,'' hep-th/9505105;\\
S. Ferrara, J. Harvey, A. Strominger and C. Vafa, ``Second Quantized
Mirror Symmetry,'' hep-th/9505162.

\bibitem{KKLMV}
S. Kachru, A. Klemm, W. Lerche, P. Mayr and C. Vafa, ``Nonperturbative
Results on the Point Particle Limit of ${\cal N}=2$ Heterotic
String Compactifications,'' hep-th/9508155.

\bibitem{SW}
N. Seiberg and E. Witten, ``Electric-magnetic duality, monopole
condensation, and confinement in ${\cal N}=2$ supersymmetric
Yang-Mills theory,'' Nucl. Phys. {\bf B426} (1994) 19, hep-th/9407087.

\bibitem{Paul}
P. Aspinwall, ``Aspects of the Hypermultiplet Moduli Space in String
Duality,'' hep-th/9802194;\\
P. Aspinwall and M.R. Plesser, ``Heterotic String Corrections from
the Dual Type II String,'' hep-th/9910248.


\bibitem{PS}
J. Polchinski and A. Strominger, ``New vacua for type II string
theory,'' hep-th/9510227.

\bibitem{Michelson}
J. Michelson, ``Compactifications of type IIB strings to
four dimensions with nontrivial classical potential,''
hep-th/9610151.


\bibitem{louismicu}
J. Louis and A. Micu, ``Type II theories compactified on Calabi-Yau
threefolds in the presence of background fluxes,'' Nucl. Phys.
{\bf B635} (2002) 395, hep-th/0202168.





\bibitem{gauging}
G.~Dall'Agata,
``Type IIB supergravity compactified on a Calabi-Yau manifold with
H-fluxes,''
JHEP {\bf 0111}, 005 (2001),
hep-th/0107264;\\
R. D'Auria, S. Ferrara and S. Vaula, ``${\cal N}=4$
gauged supergravity and a IIB orientifold with fluxes,'' New J.
Phys. {\bf 4} (2002) 71, hep-th/0206241;\\
S. Ferrara and M.
Porrati, ``${\cal N}=1$ no-scale supergravity from IIB
orientifolds,'' Phys. Lett. {\bf B545} (2002) 411,
hep-th/0207135;\\
R. D'Auria, S. Ferrara, M. Lledo and S. Vaula,
``No-scale ${\cal N}=4$ supergravity coupled to Yang-Mills: the
scalar potential and super-Higgs effect,'' Phys. Lett. {\bf B557}
(2003) 278, hep-th/0211027;\\
S.~Gurrieri, J.~Louis, A.~Micu and D.~Waldram,
``Mirror symmetry in generalized Calabi-Yau compactifications,''
Nucl.\ Phys.\ B {\bf 654}, 61 (2003), hep-th/0211102;\\
R. D'Auria, S. Ferrara, F.
Gargiulo, M. Trigiante and S. Vaula, ``${\cal N}=4$ supergravity
Lagrangian for type IIB on $T^6/Z_2$ orientifold in presence of
fluxes and D3 branes,'' JHEP {\bf 0306} (2003) 045,
hep-th/0303049;\\
L. Andrianopoli, S. Ferrara and M. Trigiante,
``Fluxes, supersymmetry breaking and gauged supergravity,''
hep-th/0307139;\\
B. de Wit, H. Samtleben and M. Trigiante,
``Maximal Supergravity from IIB Flux Compactifications,''
hep-th/0311224.

\bibitem{Kounnas}
J.P. Derendinger, C. Kounnas and F. Zwirner, ``Potentials and
superpotentials in the effective ${\cal N}=1$ supergravities
from higher dimensions,'' Nucl. Phys. {\bf B691} (2004) 233,
hep-th/0403043.


\bibitem{Acharya}
B.S. Acharya, ``A Moduli Fixing Mechanism in M-theory,''
hep-th/0212294;\\
B.S. Acharya, F. Denef, C. Hofman and N. Lambert, ``Freund-Rubin
revisited,'' hep-th/0308046;\\
B. de Carlos, A. Lukas and S. Morris, ``Non-perturbative vacua
for M-theory on $G_2$ manifolds,'' hep-th/0409255.

\bibitem{behrndt}
K.~Behrndt and M.~Cvetic,
``General ${\cal N} = 1$ supersymmetric flux vacua of (massive) type IIA string
theory,''
arXiv:hep-th/0403049;\\
K.~Behrndt and M.~Cvetic,
``General ${\cal N} = 1$ 
supersymmetric fluxes in massive type IIA string theory,''
arXiv:hep-th/0407263.


\bibitem{painful}
See for instance: L. Andrianopoli et al, ``${\cal N}=2$ Supergravity and ${\cal N}=2$
Super Yang-Mills Theory on General Scalar Manifolds: Symplectic
Covariance, Gaugings and the Momentum Map,'' hep-th/9605032.

\bibitem{Witten:1996md}
E.~Witten,
``On flux quantization in M-theory and the effective action,''
J.\ Geom.\ Phys.\  {\bf 22}, 1 (1997)
[arXiv:hep-th/9609122].

\bibitem{Antoniadis:1997eg}
I.~Antoniadis, S.~Ferrara, R.~Minasian and K.~S.~Narain,
``R**4 couplings in M- and type II theories on Calabi-Yau spaces,''
Nucl.\ Phys.\ B {\bf 507}, 571 (1997), hep-th/9707013.

\bibitem{Andy}
A. Strominger, ``Loop corrections to the universal hypermultiplet,''
Phys. Lett. {\bf B421} (1998) 139.

\bibitem{Antoniadis:2003sw}
I.~Antoniadis, R.~Minasian, S.~Theisen and P.~Vanhove,
``String loop corrections to the universal hypermultiplet,''
Class.\ Quant.\ Grav.\  {\bf 20}, 5079 (2003),
hep-th/0307268.

\bibitem{GP}
B. Greene and R. Plesser, ``Duality in Calabi-Yau moduli space,''
Nucl. Phys. {\bf B338} (1990) 15.


\bibitem{Klemm:1992tx}
A.~Klemm and S.~Theisen,
``Considerations of one modulus Calabi-Yau compactifications: Picard-Fuchs
equations, Kahler potentials and mirror maps,''
Nucl.\ Phys.\ B {\bf 389}, 153 (1993),
hep-th/9205041.




\bibitem{IIBstat}
S. Ashok and M. Douglas, ``Counting flux vacua,'' hep-th/0307049;\\
F. Denef and M. Douglas, ``Distributions of flux vacua,''
hep-th/0404116;\\
A. Giryavets, S. Kachru and P. Tripathy, ``On the taxonomy of flux
vacua,'' hep-th/0404243;\\
J. Conlon and F. Quevedo, ``On the explicit construction and
statistics of Calabi-Yau flux vacua,'' hep-th/0409215;\\
J. Kumar and J.D. Wells, ``Landscape cartography: A coarse survey of
gauge group rank and stabilization of the proton,'' hep-th/0409218;\\
O. DeWolfe, A. Giryavets, S.  Kachru and W. Taylor, ``Enumerating
flux vacua with enhanced symmetries,'' hep-th/0411061;\\ 
R. Blumenhagen, F. Gmeiner, G. Honecker, D. L\"ust and
T. Weigand, ``The statistics of supersymmetric D-brane models,''
hep-th/0411173;\\ 
F. Denef and M. Douglas, ``Distributions of nonsupersymmetric
flux vacua,'' hep-th/0411183. 




\bibitem{realistic}
D. L\"ust, S. Reffert and S. Stieberger, ``MSSM with
soft SUSY breaking terms from D7-branes with fluxes,'' hep-th/0410074;\\
D. L\"ust, S. Reffert and S. Stieberger, ``Flux induced soft supersymmetry
breaking in chiral type IIB orientifolds with D3/D7 branes,'' hep-th/0406092;\\
J. Cascales, M. Garcia de Moral, F. Quevedo and A.
Uranga, ``Realistic D-brane Models on Warped Throats: Fluxes,
Hierarchies and Moduli Stabilization,'' hep-th/0312051;\\
J. Cascales and A. Uranga, ``Chiral 4d String Vacua with D-branes and
Moduli Stabilization,'' JHEP {\bf 0402} (2004) 031, hep-th/0311250;\\
J. Cascales and A. Uranga, ``Chiral 4d ${\cal N}=1$ String Vacua with
D-Branes and
NSNS and RR Fluxes,'' JHEP {\bf 0305} (2003) 011,
hep-th/0303024;\\
R. Blumenhagen, D. L\"ust and T. Taylor, ``Moduli Stabilization in Chiral
Type IIB Orientifold Models with Fluxes,'' Nucl. Phys. {\bf B663} (2003) 219,
hep-th/0303016.

\bibitem{softsusy}
L.E. Ibanez, ``The fluxed MSSM,'' hep-ph/0408064;\\
P.G. Camara, L.E. Ibanez and A.M. Uranga, ``Flux-induced SUSY
breaking soft terms on D7-D3 brane systems,''
hep-th/0408036;\\
P.G. Camara, L.E. Ibanez and A.M Uranga, ``Flux Induced
SUSY Breaking Soft Terms, hep-th/0311241;\\
M. Grana, T. Grimm, H. Jockers and J. Louis, ``Soft Supersymmetry
Breaking in Calabi-Yau Orientifolds with D-branes and Fluxes,''
hep-th/0312232;\\
A. Lawrence and J. McGreevy, ``Local String Models of Soft Supersymmetry
Breaking,'' hep-th/0401034.





\bibitem{shiu}
F. Marchesano, G. Shiu and L. Wang, ``Model building and
phenomenology of flux-induced supersymmetry breaking on
D3-branes,'' hep-th/0411080;\\
F. Marchesano and G. Shiu, ``Building MSSM flux vacua,''
hep-th/0409132;\\
F. Marchesano and G. Shiu, ``MSSM vacua from flux compactifications,''
hep-th/0408059.

\bibitem{cvetic}
M. Cvetic and T. Li, ``Supersymmetric standard models, flux
compactification and moduli stabilization,'' hep-th/0409032;\\
M. Cvetic and T. Li, ``Supersymmetric Pati-Salam models
from intersecting D6 branes: A road to the Standard Model,''
Nucl. Phys. {\bf B698} (2004) 163;\\
M. Cvetic and I. Papadimitriou, ``More supersymmetric standard-like
models from intersecting D6-branes in type IIA orientifolds,''
Phys. Rev. {\bf D67} (2003) 126006;\\
M. Cvetic, I. Papadimitriou and G. Shiu, ``Supersymmetric
three-family $SU(5)$ grand unified models from type IIA
orientifolds with intersecting D6 branes,''
Nucl. Phys. {\bf B659} (2003) 193,
hep-th/0212177.

\bibitem{blumenhagen}
R. Blumenhagen, D. L\"ust and S. Stieberger, ``Gauge unification
in supersymmetric intersecting brane worlds,'' JHEP {\bf 0307} (2003)
036, hep-th/0305146;\\
R. Blumenhagen, L. G\"orlich and T. Ott, ``Supersymmetric
intersecting branes on the type IIA $T^6/Z_4$ orientifold,''
JHEP {\bf 0301} (2003) 021, hep-th/0211059.

\bibitem{beasleywitten}
C.~Beasley and E.~Witten, ``A note on fluxes and superpotentials in M-theory compactifications on manifolds of G(2) holonomy,''
JHEP {\bf 0207}, 046 (2002),
hep-th/0203061.




\bibitem{boussopolchinski}
R.~Bousso and J.~Polchinski,
``Quantization of four-form fluxes and dynamical neutralization of the
cosmological constant,''
JHEP {\bf 0006}, 006 (2000),
hep-th/0004134.

\bibitem{Coleman}
S.~R.~Coleman,
``More About The Massive Schwinger Model,''
Annals Phys.\  {\bf 101}, 239 (1976).


\bibitem{Cecotti:1988qn}
S.~Cecotti, S.~Ferrara and L.~Girardello,
``Geometry Of Type II Superstrings And The Moduli Of Superconformal Field
Theories,''
Int.\ J.\ Mod.\ Phys.\ A {\bf 4}, 2475 (1989).




\end{thebibliography}
\end{document}